\documentclass[%
 reprint,
 amsmath,amssymb,
 aps,
]{revtex4-2}

\usepackage{graphicx}
\usepackage{dcolumn}
\usepackage{bm}

\begin{document}

\preprint{APS/123-QED}

\title{The GOE ensemble for quasiperiodic tilings without unfolding: $r$-value statistics}

\author{Uwe Grimm}%
\email{uwe.grimm@open.ac.uk}
\affiliation{%
 School of Mathematics and Statistics, The Open University, Walton Hall, Milton Keynes, MK7 6AA, UK
}
\author{Rudolf A.\ R\"{o}mer}
\email{r.roemer@warwick.ac.uk}
\affiliation{Department of Physics, University of Warwick, Coventry, CV4 7AL, UK}


\date{\today}

\begin{abstract}
We study the level-spacing statistics for non-interacting Hamiltonians defined on the two-dimensional quasiperiodic Ammann--Beenker (AB) tiling. When applying the numerical procedure of ``unfolding'', these spectral properties in each irreducible sector are known to be well-described by the universal Gaussian orthogonal random matrix ensemble. However, the validity and numerical stability of the unfolding procedure has occasionally been questioned due to the fractal self-similarity in the density of states for such quasiperiodic systems. Here, using the so-called $r$-value statistics for random matrices, $P(r)$, for which no unfolding is needed, we show that the Gaussian orthogonal ensemble again emerges as the most convincing level statistics for each irreducible sector. The results are extended to random-AB tilings where random flips of vertex connections lead to the irreducibility.
\end{abstract}

\maketitle





The statistical description of energy levels in complex systems has a long and distinguished history \cite{Bethe1936AnNucleus,Wigner1951OnLevels}. For interacting systems such as, e.g., highly excited heavy nuclei, already Bethe recognized the intrinsic difficulty in obtaining such statistics \cite{Bethe1936AnNucleus} where the spacing of levels is often complicated and the Hilbert space exponentially large. Much progress has been made, nevertheless, when ignoring details of the underlying Hamiltonian and instead assuming a random matrix structure \cite{Dyson1962StatisticalI,Dyson1962StatisticalII,Dyson1962StatisticalIII,Dyson1963StatisticalIV,Metha}. In such a situation, it is the invariance of the matrix under specific symmetry operations that determines the functional form of the level spacing distribution $P(s)$, with Gaussian orthogonal, unitary and symplectic ensembles, GOE, GUE, GSE, respectively, being the most famous examples \cite{Metha}. 


Difficulties in using level statistics also arise for quasiperiodic (QP) systems when electronic degrees of freedom are described by non-interacting tight-binding Hamiltonians defined on QP tilings. It is well known that for such systems, the density of states (DOS) typically has self-similar (fractal) characteristics while a straightforward comparison with the Gaussian ensembles requires a flat DOS. A procedure known as ``unfolding'' is often used to convert the fractal DOS into the required flat behaviour but this of course represents a severe change in the energy spectrum. Indeed, it has been shown that $P(s)$ can change when only partial unfolding is being done \cite{Jagannathan2007EnergyQuasicrystals}. Using the integrated DOS instead as basis for an unfolding procedure yields more stable results which demonstrate that $P(s)$ for a time-invariant and spin-independent QP Hamiltonian is very well described by $P_\text{GOE}(s)$ \cite{ZHONG1998}. In fact, the $P(s)$ was shown to follow $P_\text{GOE}(s)$ better than the celebrated Wigner expression $P_\text{Wigner}(s)= \pi/ 2 \ \exp{(- \pi s^2 /4)}$ of GOE which is based on a $(2 {\times} 2)$-matrix surmise \cite{Wigner1951OnLevels}.


Removing the dependence on the unfolding procedure, Oganesyan and Huse \cite{Oganesyan2007a} introduced the so-called $r$-value distribution in the context of disordered many-body systems. With $s_n= E_n - E_{n-1}$ denoting the spacing of two consecutive energy levels $E_n$, $E_{n-1}$, they define $r_n$ as
\begin{equation}
    0 \leq r_n = \text{min}\{s_n, s_{n-1} \} / \text{max}\{s_n, s_{n-1} \} \leq 1 .
\end{equation}
For an uncorrelated Poisson spectrum, one has $P_\text{Poisson} = 2/(1+r)^2$ with mean $\langle r \rangle_\text{Poisson} = 0.386$ while for the GOE, 
\begin{equation}
    P_\text{GOE}(r) \sim \frac{27 (r + r^2) }{4(1 + r + r^2)^{5/2}} 
    \label{eq:PGOEr}
\end{equation}
with $\langle r \rangle_\text{GOE} \sim 0.5307(1)$ \cite{Atas2013DistributionEnsembles}. Expression \eqref{eq:PGOEr} has the same status as the Wigner surmise quoted above, i.e., has been derived for the smallest possible ($3 {\times} 3$) matrix in the GOE while $\langle r \rangle_\text{GOE}$ is based on high-precision numerics. Here, we shall also use a surmise based on a $(5 {\times} 5)$-matrix (see appendix) which improves upon \eqref{eq:PGOEr} such that the deviation to $\langle r \rangle_\text{GOE}$ is reduced from $1\%$ to $< 0.4\%$. In the context of many-body localization (MBL), $r$-value statistics has proven its worth by allowing the numerical determination of the transition from MBL to the so-called ergodic phase at weak disorders without the need to unfold spectra \cite{Oganesyan2007a,Nandkishore2014b,Abanin2019Colloquium:Entanglement}. 


\begin{figure*}[tb]
    \centering
    (a)\includegraphics[width=0.95\columnwidth]{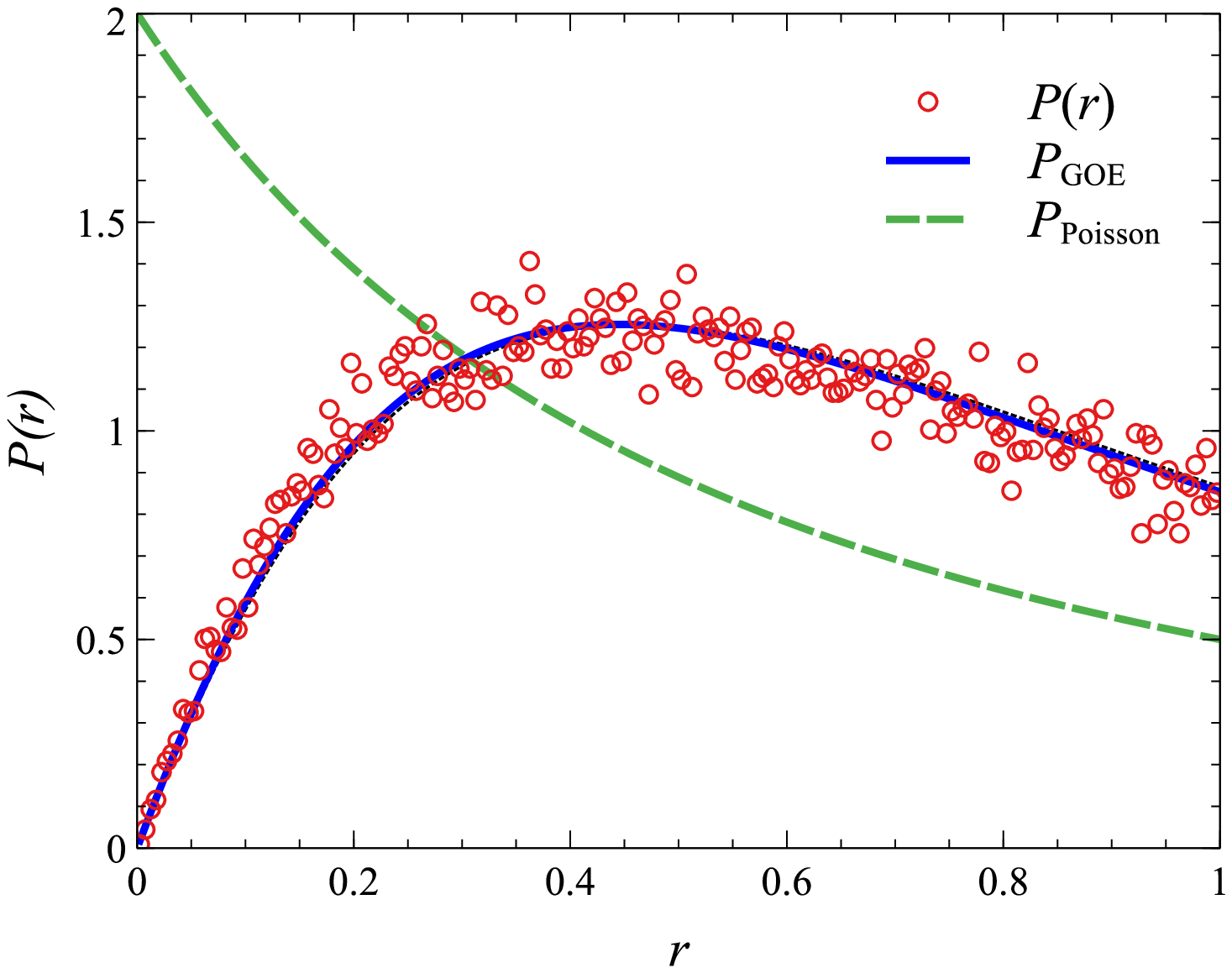}
    (b)\includegraphics[width=0.95\columnwidth]{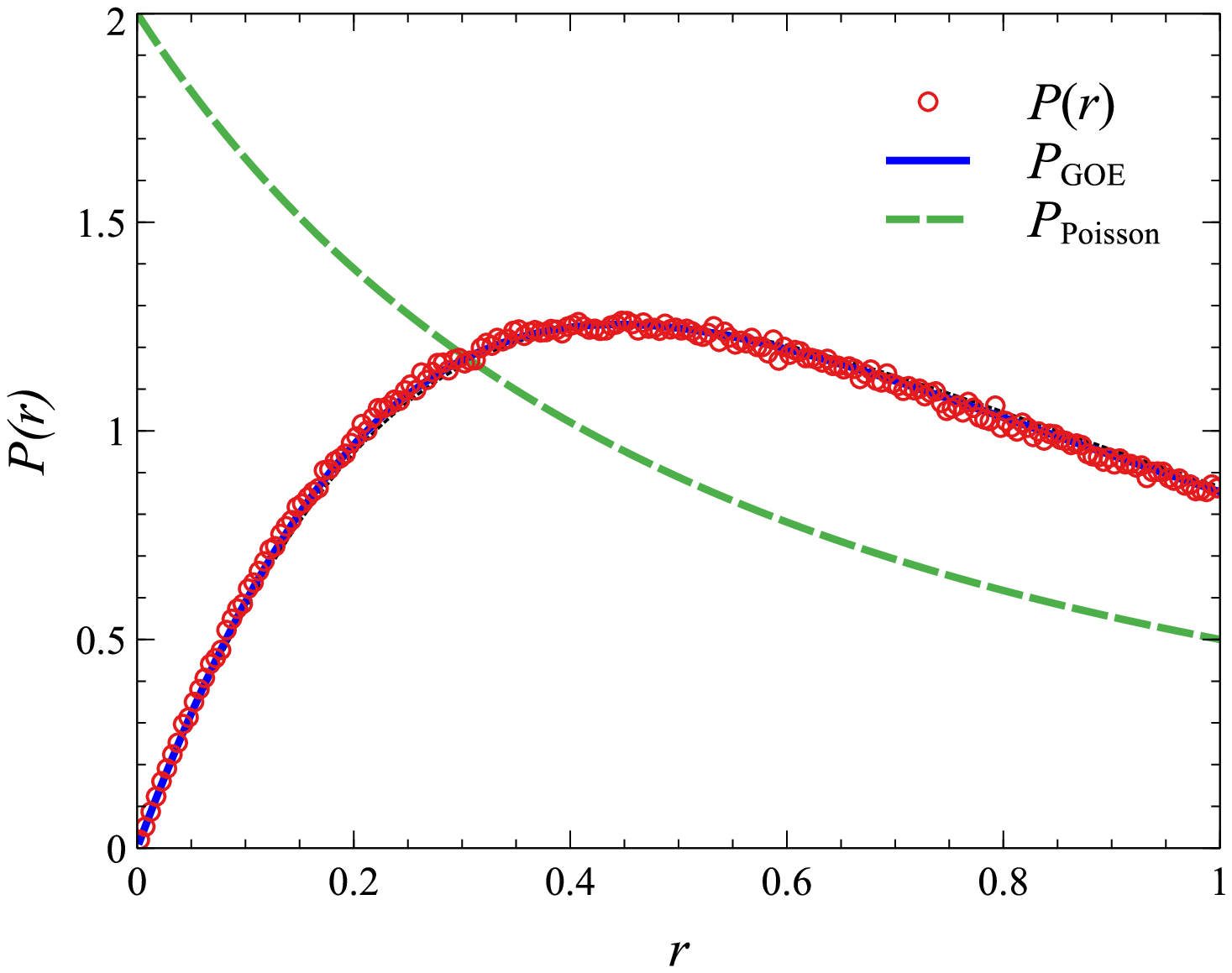}
    \caption{Plot of $P(r)$ for (a) the combined $P(r)$ (red dots) for all seven sectors of the AB tiling with inflation step $\mathcal{I}=5$ corresponding to $N=157369$ vertices and (b) the combined $P(r)$ (red dots) for the $300$ (combined) realizations of the largest random-AB tiling at $\mathcal{I}=4$ as detailed in Table \ref{tab:rvalues} . In both panels, the dashed (green) lines gives $P_\text{Poisson}(r)$ while the solid (blue) lines represent the improved $P_\text{GOE}(r)$ and the dotted (black) line is \eqref{eq:PGOEr}. 
    }
    \label{fig:PGOEr}
\end{figure*}

In this paper, we employ the $r$-value statistics to QP Hamiltonians defined on the Ammann--Beenker (AB) octagonal tiling \cite{Ammann1992AperiodicTiles} and also to randomized versions of the tiling when certain connections have been allowed to flip. As we show in Fig.~\ref{fig:PGOEr}, in both cases, we find that the computed $P(r)$ agrees very well with the predictions of the GOE when compared to $P_\text{GOE}(r)$. Hence, within the numerical accuracy available to us, we can conclude that the GOE ensemble is indeed the correct descriptor of level statistics for non-interacting tight-binding hopping Hamiltonians in QP tilings, whether that result has been computed (i) by unfolding DOS \cite{Jagannathan2007EnergyQuasicrystals} or IDOS \cite{ZHONG1998}, (ii) by restricting the analysis to regions in the spectrum that have a flat DOS and hence do not need unfolding \cite{SCHREIBER1999b} or (iii) by circumventing unfolding with the $r$-value statistics.



\begin{figure*}[bt]
    \centering
    (a)\includegraphics[width=0.55\columnwidth]{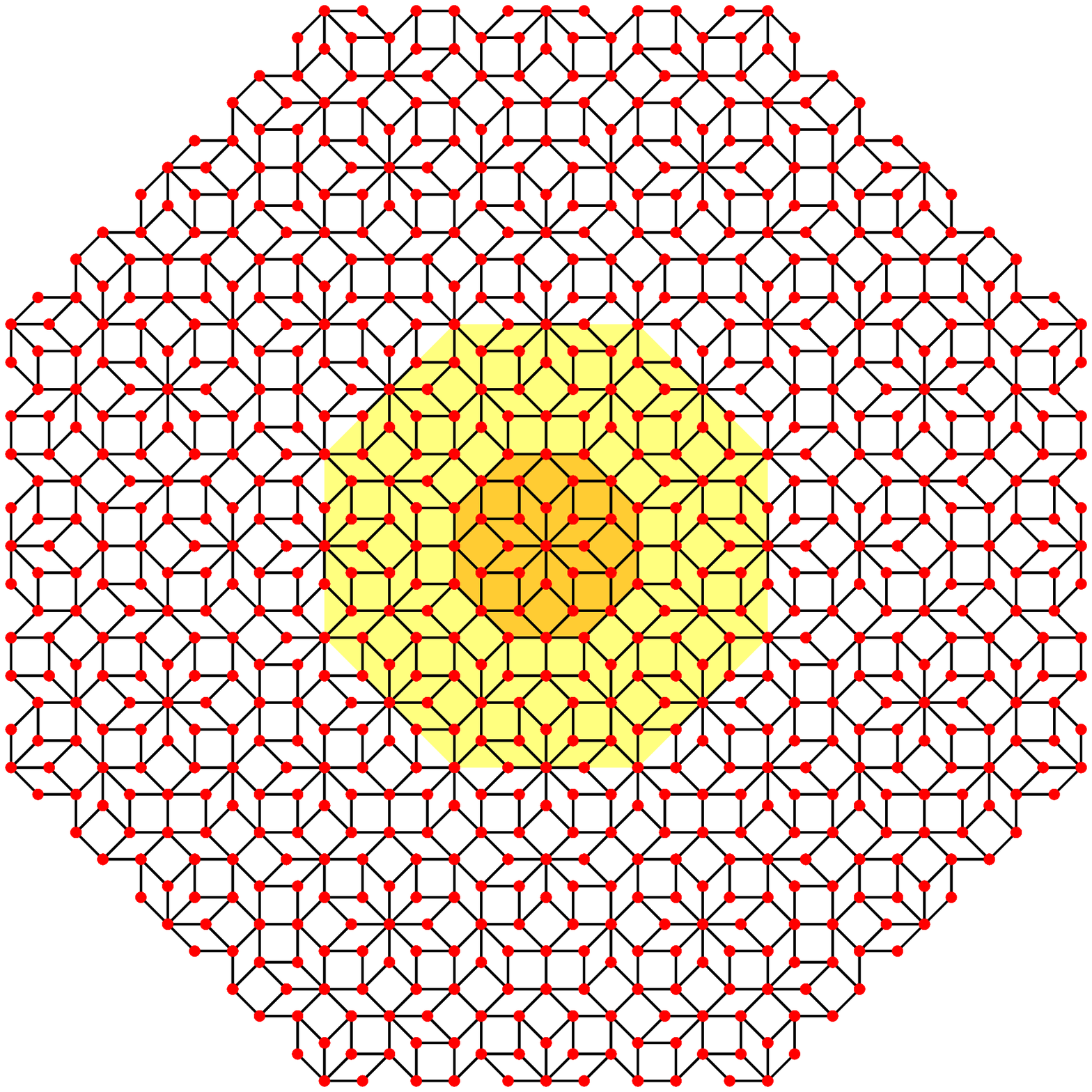}
    \includegraphics[angle=90,width=0.3\columnwidth]{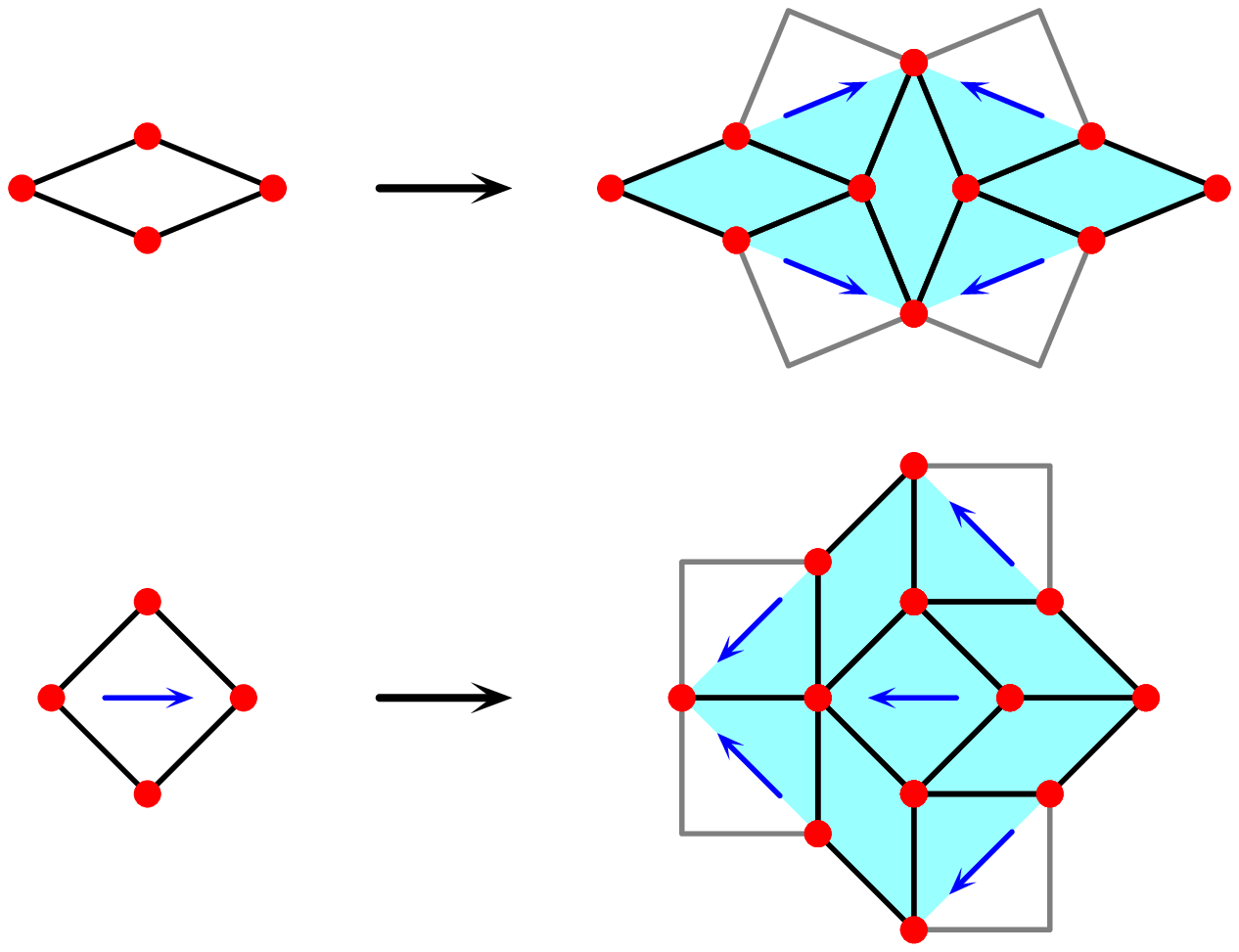}
    (b)\includegraphics[width=0.47\columnwidth]{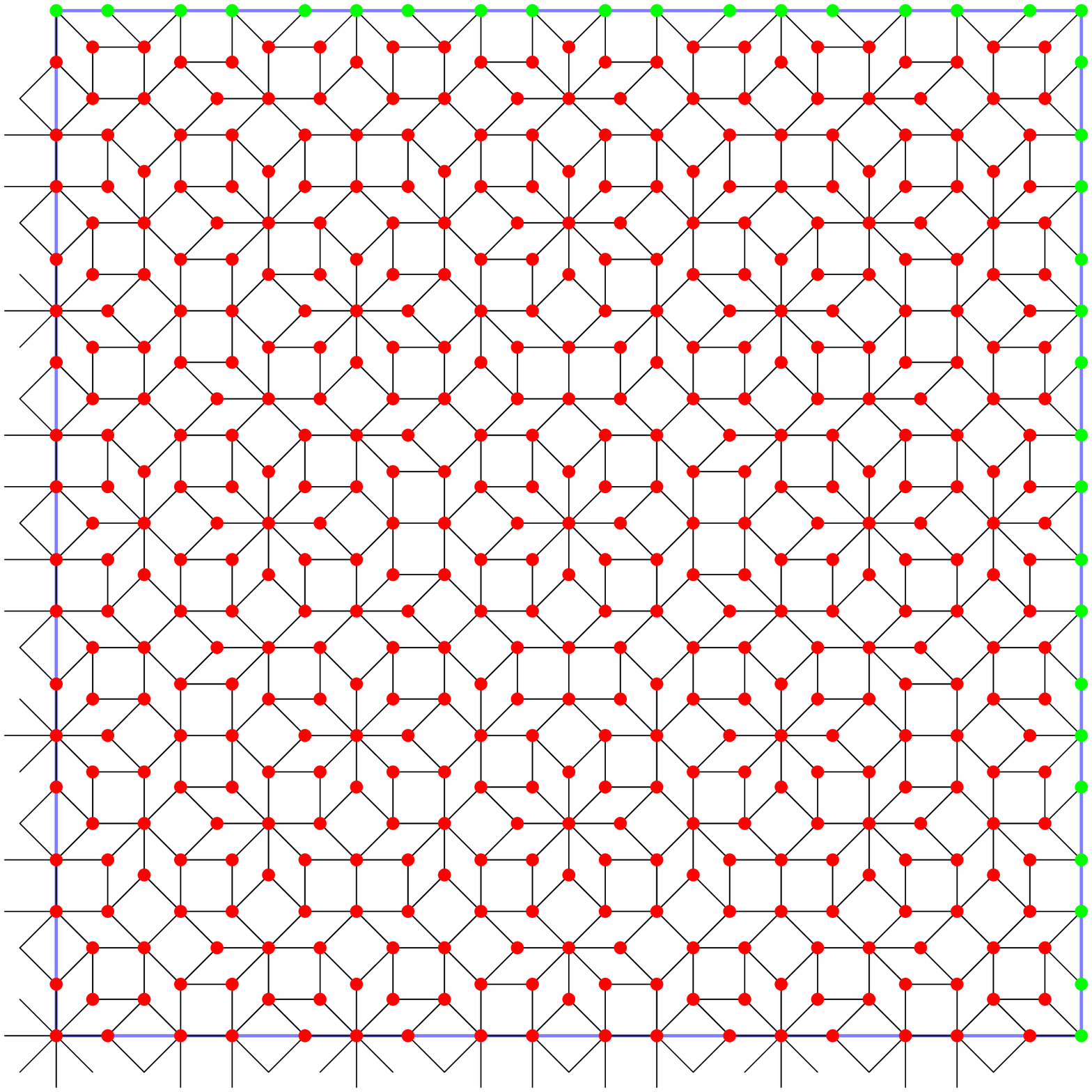}
    (c)\includegraphics[width=0.47\columnwidth]{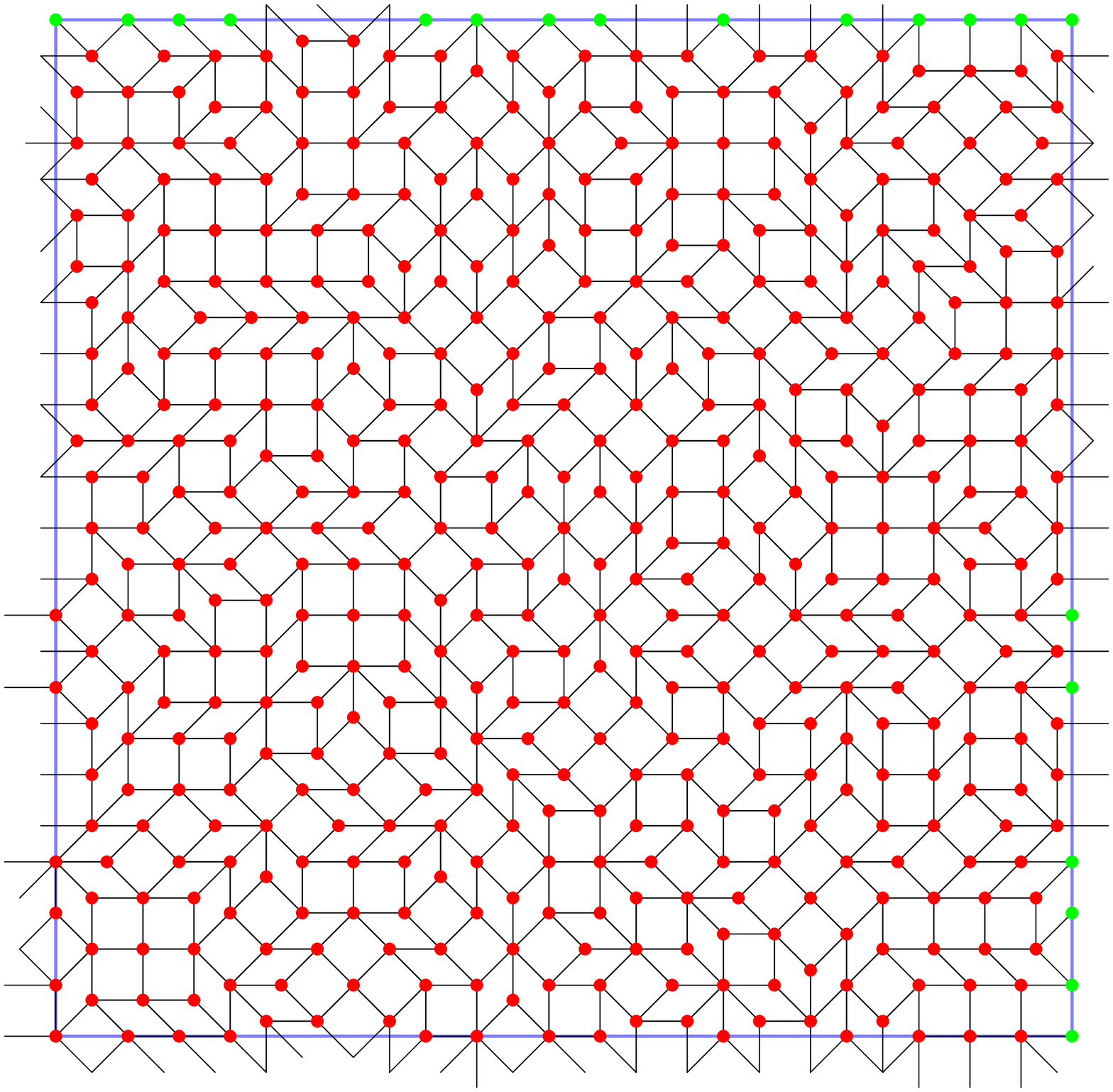}
    \includegraphics[angle=90,width=0.08\columnwidth]{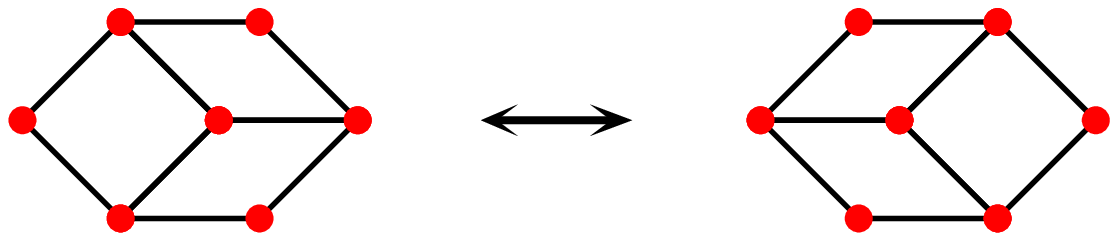}
    \caption{Schematics of (a) the octagonal AB tiling with $N=833$ vertices (red dots) and exact $D_8$ symmetry around the central vertex. Background shades/colors indicate successive inflation steps ($\mathcal{I}=0,1,2$) of the central (orange/dark) octagon. Details of the inflation procedure are given by the two small figures which show how to inflate each rhombus and square. Arrows along the diagonals of the squares provide directions that are required to define the inflation rule, since the dissection of the square breaks its symmetry. The results tiling, when starting from a symmetric seed like in part (a), keeps the $D_8$ symmetry for each $\mathcal{I}$.
    (b) A perfect periodic approximant of the AB tiling with $N=478$ and 
    (c) a random-AB tiling, also of size $N=478$. The small figure on the right shows an individual ``singelton flip'' which induces the  randomness.
    All tilings (a)-(c) correspond to $\mathcal{I}=2$. 
    The fundamental regions for (b) and (c) are enclosed within the blue lines, forming a square. The green dots on the blue line denote the periodically replicated vertices for periodic boundary conditions. The thin solid lines connecting vertices $i$, $j$ indicate the neighbor connection $\langle i,j \rangle$ as used in the Hamiltonian.
    }
    \label{fig:schematics}
\end{figure*}

As in Ref.~\onlinecite{ZHONG1998}, we shall consider the  octagonal (or Ammann--Beenker) \cite{Ammann1992AperiodicTiles,Duneau1989ApproximantsMapping} tiling consisting of squares and rhombi with equal edge lengths as in Fig.~\ref{fig:schematics}(a); see \cite{BaakeGrimm2013} for more on this tiling and its properties. 
Besides this perfect quasicrystal, we shall also study a randomized version in which triangular connections in the rhombi are allowed to flip randomly. Such structures are often used to model imperfect quasicrystals; see e.g.~\cite{HenElsMih2000}. Both the AB tilings and the random-AB tilings increase in size exponentially via inflation as $N \sim (3+2\sqrt{2})^{\mathcal{I}}=(1+\sqrt{2})^{2\mathcal{I}}$ for $\mathcal{I}\rightarrow\infty$, where $\mathcal{I}\in \mathbb{N}$ denotes the generation of the inflation with $\mathcal{I}=0$ corresponding to the initial patch (the seed). We include data for patches of the AB tiling corresponding to $\mathcal{I}=2$, $3$, $4$, and $5$ inflation steps with $833$, $4713$, $27137$, and $157369$ vertices, respectively. For the random-AB tiling, we use $\mathcal{I}=1$, $2$, $3$ and $4$ inflation steps with $82$, $478$, $2786$ and $16238$ vertices, respectively.
On these tilings, we define $H= \sum_{\langle i,j \rangle} |i\rangle \langle j|$ as Hamiltonian with free boundary conditions for the AB tilings and periodic boundary conditions for the random-AB tilings. Here, $| i \rangle$ is indicating the Wannier state at vertex $i$ while pairs of neighboring vertices connected by an edge of unit length are denoted as $\langle i, j \rangle$.

The AB tiling in Fig.~\ref{fig:schematics}(a) has the symmetry of the regular octagon, corresponding to the dihedral group $D_8$. Hence the Hamiltonian matrix splits into ten blocks according to the irreducible representations of $D_8$: using the rotational symmetry, one obtains eight blocks, two of which split further under reflection, while the remaining six form three pairs with identical spectra. This gives a total of seven independent subspectra. 
The starting point in the generation of the random-AB tilings is the perfect periodic approximant shown in Fig.\ \ref{fig:schematics}(b). Note that this is the periodic approximant of highest exact symmetry for this tiling, since eightfold symmetry is forbidden in a periodic structure. As such, it has $D_4$ symmetry and five independent spectra. 
We can introduce randomness by flipping the arrangement of hexagonal patches consisting of a square on two rhombi that meet in a three-valent vertex. These ``simpleton flips'' are ergodic in the sense that repeated applications of this flip explore the entire ensemble of random tilings of square and rhombi with the given ratio of the two tiles. We shall study cases with an increasing number of flips per vertex. Note that the flips generally will remove any exact symmetries, such that the whole matrix becomes an irreducible block, while the statistical eightfold symmetry (in the sense that local configurations are equally likely to appear in any of the eight directions) is maintained; see \cite{BaakeGrimm2013} for details. As example is given in Fig.~\ref{fig:schematics}(c).
For both AB tilings and random-AB tilings, the Hamiltonian is diagonalized independently for each of the irreducible spectra; each spectrum is symmetric about $E=0$, because of the bipartiteness of the AB and random-AB tilings. Furthermore, a finite fraction of the states is degenerate at $E=0$, corresponding to compactly-localized states \cite{Sutherland1986LocalizationTopology,Kohmoto1986,Leykam2018ArtificialExperiments}; they do not contribute to the universal statistics, and we neglect them.

As is well known, the DOS for each irreducible spectrum is rather spiky \cite{Jagannathan2007EnergyQuasicrystals}, while the IDOS is already rather smooth \cite{ZHONG1998} (results not shown here). 
We proceed without any unfolding.
Only eigenvalues $|E_n|>10^{-10}$ are included in the further analysis and in Table~\ref{tab:rvalues}, we give the number of these as $N(E\neq 0)$. A further restriction to positive $E_n$'s also removes the double degeneracy, for the $s_n$, resulting from the bipartiteness of the tilings. This leads to the available number of $r_n$-values given in Table~\ref{tab:rvalues} as $N(r)$. For each spectrum, we independently compute $P(r)$ and $\langle r \rangle$. 
For the AB tilings, we find that the $P(r)$ distributions for all $\mathcal{I}$ show level repulsion such that $P(r) \propto r$ for $r \rightarrow 0$ \cite{Atas2013DistributionEnsembles}. With increasing $\mathcal{I}$, the slope of the level repulsion decreases slightly, level repulsion increases and rapidly approaches the small-$r$ behaviour of $P_\text{GOE}(r)$. Similarly, the bulk behaviour of $P(r)$ follows $P_\text{GOE}(r)$ ever more closely for increasing $\mathcal{I}$. 
In order to ascertain quantitatively how well the estimated $P(r)$ follows either $P_\text{Poisson}(r)$ or $P_\text{GOE}(r)$, we establish the root-mean-squared deviation (RMSD) defined as $\{ \int_{0}^{\infty} [P(r) - Q(r) ]^2 \text{d}r \}^{1/2}$ with $Q(r)$ either $P_\text{Poisson}(r)$ or $P_\text{GOE}(r)$. In Table \ref{tab:rvalues} we also express the RMSD values in percentage by comparing with the RMSD$_\text{max}= \{ \int_{0}^{\infty} [ P_\text{Poisson}(r) -P_\text{GOE}(r)]^2 \text{d}r \}^{1/2} = 0.630508$ between $P_\text{Poisson}(r)$ and $P_\text{GOE}(r)$. We find for $P(r)$ that while the RMSD to GOE values, RMSD$_\text{GOE}$, are roughly in the $\sim 10-20\%$ range, the RMSD$_\text{Poisson}$ values are at $\sim 90-100\%$ of RMSD$_\text{max}$. Hence for each subspectrum, we see that $P(r)$ already very nicely follows $P_\text{GOE}(r)$ while $P_\text{Poisson}(r)$ is certainly ruled out.
This conclusion is also corroborated when studying $\langle r \rangle$ for the AB tilings with all estimates within $2\%$ of $\langle r \rangle_\text{GOE} \sim 0.5307(1)$.
The best agreement is found when we combine all $r$-values for the largest system $\mathcal{I}=5$. The resulting $P(r)$ is the one shown in Fig.~\ref{fig:PGOEr} with $\langle r \rangle = 0.526654(1)$.

Due to the exponential growth of the size of the Hamiltonian matrix with $\mathcal{I}$, a further increase of $\mathcal{I}$ is computationally very challenging (for $\mathcal{I}=6$, we have $N=1,657,756,990$) \cite{I6}. We therefore now turn to the random-AB tilings introduced above where we can increase $N(r)$ simply by computing many random realizations. In these systems, the DOS is also somewhat less spiky, but still retains considerable variation across the energy spectrum that would still require significant unfolding when studying $P(s)$. 
We summarize the results in Table \ref{tab:rvalues} and show the behaviour of $P(r)$ in Fig.~\ref{fig:PGOEr}. For $\mathcal{I}=3$, we give results for all $1000$ samples when flipping each triangular connection (cf.\ Fig.~\ref{fig:schematics}(c)) on average $1$, $10$, $100$ or $1000$ times for a thoroughly randomized tiling. We find that the differences in $P(r)$ between these cases, even with just a single flip per vertex (on average), are very small, and no major influence of the underlying exact $D_4$-symmetry of the approximant can be seen anymore. We therefore also present ``combined'' statistics where $r$-values for $10$, $100$ and $1000$ flips have been analyzed together.
For $\mathcal{I}=2$ we only show these combined results while full details are given also for $\mathcal{I}=4$. In this case, the computational effort is already considerable for each sample so that only $100$ samples have been calculated for all flips. The overall combined $N(r)=699579$, $4132302$ and $2421369$ for $\mathcal{I}= 2, 3$ and $4$, respectively, are already considerable larger than the $N(r)= 45049$ available for $\mathcal{I}=5$ in the case of the AB tilings.
With this increased statistical sample, we find that the agreement with $P_\text{GOE}(r)$ is now even better, particularly for $\mathcal{I}=4$. The final $\langle r\rangle= 0.530347$ value is indeed within $\sim 0.07 \%$ of $\langle r \rangle_\text{GOE}$.

\begin{table*}{}
\centering\setlength{\tabcolsep}{6pt}
\begin{tabular}{clrrrllclc}
\hline \hline 
$\mathcal{I}$\rule[-1.5ex]{0ex}{4.5ex} & sector/flips &   $N(E\neq 0)$&  parts/samples & $N(r)$ & $\langle r \rangle$ & RMSD$_\text{GOE}$ & \% & RMSD$_\text{Poisson}$ & \% \\ \hline

\vspace*{.1ex} \\
\multicolumn{10}{c}{AB tilings \cite{I6}} \\[1ex]

6 & 0 (-1)	& 56833	& -- & 25989	&0.52408(1) 	& 0.081306	& 13	&0.605505	& 97 \\	
6 & 0 (1)	& 57580	& -- & 26366	&0.52554(1)	    & 0.076748	& 12	&0.605635	& 97 \\
6 & 4 (-1)	& 57241	& -- & 26196	&0.52354(1)	    & 0.086207	& 14	&0.605097	& 97 \\
6 & 4 (1)	& 57171	& -- & 26156	&0.52848(1)     & 0.074022	& 12    &0.119028	& 98 \\
6 & combined & 228825 &	4 &	104707	&\textbf{0.525417(3)}    & 0.050434	& 8     &0.081114	& 97 \\[1ex]

5 & 0 (-1)	& 9681  & -- &	4425 &	0.52657(6)		& 0.180536	& 29	& 0.597355	& 94 \\
5 & 0 (1)	& 9991  & -- &	4582 &	0.53139(6)		& 0.0858138	& 14	& 0.608586	& 97 \\
5 & 1	    & 19671 & -- &	9019 &	0.52327(3)		& 0.151593	& 24	& 0.611349	& 97 \\
5 & 2	    & 19671 & -- &	9019 &	0.52587(3)	    & 0.144386	& 23    & 0.603325	& 96 \\
5 & 3	    & 19671 & -- &	9019 &	0.53116(3)		& 0.134957	& 21	& 0.616345	& 98 \\
5 & 4 (-1)	& 9850  & -- &	4511 &	0.52103(6)		& 0.120176	& 19	& 0.580059	& 92 \\
5 & 4 (1)   & 9821  & -- &	4494 &	0.52684(6)		& 0.137588	& 22	& 0.607249	& 96 \\
5 & combined & 98356 & 7 &	45069 &	\textbf{0.526654(1)}		& 0.0751708    & 12	& 0.603823	& 96 \\[1ex]

 4 & combined	& 16961	& 7	& 7780	& 0.52105(3)	& 0.092353	& 15	& 0.590098	& 94\\

3 & combined    & 2946	& 7	& 1345	& 0.5323(2)	    & 0.104493	& 17	& 0.608367	& 96	\\

2 & combined	& 521	& 7	& 225	& 0.532(1)	    & 0.159398	& 25	& 0.663913	& 105 \\[1ex]


5 & mult.: 4 ($\pm 1$) & 19671	& 2	& 9007	& 0.41971(4) & 	- & - & - & - \\


5 & mult.: all 
& 98356	& 7	& 45081	& 0.391728(7)	& 	- & - & - & - \\

\vspace*{.1ex} \\
\multicolumn{10}{c}{random-AB tilings} \\[1ex]


4 & 1 	 & 1623800	& 100	& 811269	& 0.529697	& 0.0193322	& 3	& 0.615156	& 99 \\
4 & 10 	 & 1623800	& 100	& 804434	& 0.530519	& 0.0173488	& 3	& 0.61614	& 99 \\
4 & 100	 & 1623800	& 100	& 804393	& 0.530211	& 0.0178077	& 3	& 0.614561	& 99 \\
4 & 1000 & 1623800	& 100	& 812542	& 0.530312	& 0.0180712	& 3	& 0.615265	& 99 \\
4 & combined & 4887638 & 300	& 2421369	& \textbf{0.530347}	& 0.0126909	& 2	& 0.615197	& 99 \\[1ex]


3 & 1	 & 2786000	& 1000	& 1375849	& 0.521659	& 0.0458914	& 7	& 0.58396	& 93 \\
3 & 10	 & 2786000	& 1000	& 1377081	& 0.523429	& 0.0388161	& 7	& 0.590847	& 95 \\
3 & 100	 & 2786000	& 1000	& 1377642	& 0.523263	& 0.0388161	& 6	& 0.591098	& 95 \\
3 & 1000 & 2786000	& 1000	& 1377579	& 0.523954	& 0.0388161	& 6	& 0.593182	& 96 \\
3 & combined	& 8358000	& 3000	& 4132302	& 0.523548	& 0.0368666	& 6	& 0.591631	& 93 \\[1ex]


2 & combined	& 1434000	& 3000	& 699579	& 0.502911	& 0.131713	& 22	& 0.502227	& 81 \\[1ex]

\hline\hline
\end{tabular}
\caption{Table of $r$-value estimates and RMSD for (top) AB and (bottom) random-AB tilings for different inflation levels $\mathcal{I}$. The column ``sector/flips'' gives the phase (and parity) of the subspectrum for a particular AB tiling, while for the random-AB tilings, it indicates the chosen average value of flips per vertex. Labels indicate whether the statistics of different sectors were ``combined'' for $r$-values from different sectors or whether "mult"(iple) energy spectra were analyzed together.
$N(E\neq 0)$ indicates the number of non-zero energy levels, while $N(r)$ counts to number of $r$ values used to construct $P(r)$. The column ``parts/samples'' labels the number of subspectra for each AB tiling and the total number of spectra for the random-AB tilings.
The average is given by $\langle r \rangle = \int_{0}^{1} r p(r) \text{d}r$ and RMSD$_\text{GOE/Poisson}$ show the RMSD with respect to either GOE or Poisson ensemble. The error estimates for AB tilings are computed as error-of-mean; for the random-AB tilings, these are less than $10^{-6}$ and not indicated here.
The columns headed by \% show how the computed RMSD$_\text{GOE/Poisson}$ deviate from RMSD$_\text{max}$. The averages for AB tilings with combined statistics for $\mathcal{I}=5$ and for the random-AB tiling with $\mathcal{I}=4$ are highlighted in bold.}
\label{tab:rvalues}
\end{table*}


In presenting the results shown in this work, we have been careful to only show level statistics computed for spectra consisting of irreducible blocks of the Hamiltonians. If we were to not separate these irreducible sectors (according to phase and parity) we would of course get a $P(r)$ that becomes progressively closer to $P_\text{Poisson}(r)$ just as is the case for $P(s)$ statistics \cite{ZHONG1998}. 
Surmises for such spectra are only known for $P(s)$ \cite{Basor1992AsymptoticsMatrices} and not yet for $P(r)$ \cite{Atas2013DistributionEnsembles}, but reliable estimates for $\langle r \rangle$ exist \cite{Giraud2020}. In good agreement with these latter results, we find 
$\langle r \rangle = 0.41971(4)$ for $\mathcal{I}=5$, when combining parities $\pm 1$ for sector $0$ (2 irreducible blocks) as well as $\langle r \rangle = 0.391728(7)$ when using all seven sectors of the AA tiling (cf.\ Table \ref{tab:rvalues}). The results from Ref.\ \cite{Giraud2020} give $\langle r \rangle =0.423415$ and $0.391048$, respectively, for these two cases when weighting according to $N(E\geq 0)$.

Quasicrystals represent a material class between periodic crystals and aperiodic solids. As such, it has earlier been speculated that they might posses non-standard level statistics \cite{Benza1991BandChaos,Piechon1995Energy-levelQuasicrystal}. However, our results allow us to conclude that both $P(s)$ \cite{ZHONG1998} and $P(r)$ statistics for two-dimensional, QP tight-binding models are, within the numerical accuracy currently achievable, very well described by the GOE ensemble. For $P(s)$, this holds after unfolding \cite{Jagannathan2007EnergyQuasicrystals} such that even the small difference between $P_\text{GOE}(s)$ and $P_\text{Wigner}(s)$ is resolved. For $P(r)$, as we show here, also the unfolding procedure becomes superfluous to reach the same conclusion. 

\acknowledgments
We are grateful to Fabien Alet for pointing us to Ref.\ \cite{Giraud2020} and providing $\langle r\rangle$ values for combinations of irreducible blocks.
We thank Warwick's Scientific Computing Research Technology Platform for computing time and support. UG gratefully acknowledges support from EPSRC through grant EP/S010335/1. UK research data statement: Data accompanying this publication are available in \cite{Grimm2021DataStatistics}.

\appendix\section*{Appendix}
To compute an improved approximation to $P_{\text{GOE}}(r)$, we follow
Ref.~\onlinecite{Atas2013DistributionEnsembles} and perform the analogous calculation for the $(5{\times} 5)$-matrix case. The joint probability distribution for the GOE ensemble for the $(5{\times} 5)$-matrix case  
is \cite{Metha}
\[
\varrho(e_1,\dots,e_5) = C_5 \!\prod_{1\le i<j\le 5} \!\lvert e_i-e_j\rvert\,
\prod_{i=1}^{5}\exp(-e_i^2/2),
\]
where $C_5$ is the normalization constant. The distribution $P^{(5)}_\text{GOE}(r)$ 
for the $(5{\times} 5)$-matrix case can then be computed as
\[
\int\limits_{-\infty}^{\infty}\!\!\!\mathrm{d}e_3\!\!
\int\limits_{-\infty}^{e_2}\!\!\!\mathrm{d}e_1\!\!
\int\limits_{-\infty}^{e_3}\!\!\!\mathrm{d}e_2\!
\int\limits_{e_3}^{\infty}\!\!\mathrm{d}e_4\!
\int\limits_{e_4}^{\infty}\!\!\mathrm{d}e_5
\,\varrho(e_1,\dots,e_5) \,\delta\bigl(r-\tfrac{e_4-e_3}{e_3-e_2}\bigr),
\]
where we considered the eigenvalues to be ordered, with $e_1\le e_2\le e_3\le e_4\le e_5$, and concentrated
on the spacing around the central eigenvalue $e_3$, which we believe to provide a better approximation than taking the average over the three terms arising from the spacings around $e_2$, $e_3$ and $e_4$. When computing $\langle r \rangle = \int_{0}^{1} r P^{(5)}_\text{GOE}(r) \text{d}r=0.532592$, we see that the result is even closer to the high-precision numerical estimate $\langle r \rangle_\text{GOE} \sim 0.5307(1)$ than the numerical corrections using $\delta P(r)$ ($0.524912$) as proposed in Ref.\ \cite{Atas2013DistributionEnsembles}.
However, evaluating the $5$ integrals results in a lengthy expression for $P^{(5)}_\text{GOE}(r)$; for details of the computation and the result, we refer to a Mathematica notebook \cite{Grimm2021DataStatistics}. We note that a systematic study for increasing $(N {\times} N)$-matrices has been done previously in Ref.\ \cite{Atas2013JointMatrices}.


\begin{thebibliography}{26}
\bibitem{I6} After acceptance for publication, results for the 4 sectors of the AB tiling at $\mathcal{I}=6$ with additional parity symmetry became available to us. These have now been included in Table \ref{tab:rvalues}. While the estimate for $\langle r \rangle$ is slightly shifted away from $\langle r \rangle_\text{GOE} \sim 0.5307(1)$, the agreement with $P_\text{GOE}$ continues to increase.
\end{thebibliography}

%

\end{document}